\documentclass[epl,twocolumn,showpacs,preprintnumbers,amsmath,amssymb]{revtex4}
\usepackage{graphicx}
\usepackage{dcolumn}
\usepackage{bm}

\begin{document}

\title{Design of a low band gap oxide ferroelectric:
Bi$_6$Ti$_4$O$_{17}$}

\author{Bo Xu}
\affiliation{Physics Department, National University of Singapore,
2 Science Drive 3, Singapore 117542, Singapore}
\affiliation{College of Physics and Communication Electronics,
Jiangxi Normal University, Nanchang, Jiangxi,
People's Republic of China 33022}

\author{David J. Singh}
\author{Valentino R. Cooper}
\affiliation{Materials Science and Technology Division,
Oak Ridge National Laboratory, Oak Ridge, TN 37831-6114, USA}

\author{Yuan Ping Feng}
\affiliation{Physics Department, National University of Singapore,
Singapore 117542, Singapore}

\date{\today}

\begin{abstract}
A strategy for obtaining low band
gap oxide ferroelectrics based on charge imbalance is described
and illustrated by
first principles studies of the
hypothetical compound Bi$_6$Ti$_4$O$_{17}$, which is an
alternate stacking of the ferroelectric Bi$_4$Ti$_3$O$_{12}$.
We find that this compound is ferroelectric, similar to Bi$_4$Ti$_3$O$_{12}$
although with a reduced polarization. Importantly, calculations of the
electronic structure with the recently developed functional of Tran and Blaha
yield a much reduced band gap of 1.83 eV for
this material compared to Bi$_4$Ti$_3$O$_{12}$.
Therefore, Bi$_6$Ti$_4$O$_{17}$ is predicted
to be a low band gap ferroelectric material.
\end{abstract}

\pacs{78.20.Ci,77.55.fp,71.20.Ps,71.15.Mb}

\maketitle

Oxide ferroelectrics suitable for applications generally have
band gaps of 3 eV or higher.
However, observations
of interesting photovoltaic effects that may be of practical
importance
\cite{choi,ichiki,pintilie,yang,huang}
have led to interest in materials with
stronger light absorption and lower band gaps.
\cite{bennett}
In this regard, Ti is a particularly interesting element.
First of all it occurs in a variety of well known useful ferroelectric
materials, such as BaTiO$_3$, PbTiO$_3$ and Bi$_4$Ti$_3$O$_{12}$.
Secondly,
it readily forms oxides in two valence states,
Ti$^{4+}$ (e.g. TiO$_2$ and
BaTiO$_3$) and Ti$^{3+}$ (e.g. Ti$_2$O$_3$ and LaTiO$_3$).
Finally, interfaces of SrTiO$_3$ with LaTiO$_3$ or LaAlO$_3$ have been
shown to be metallic with high mobility. \cite{ohtomo}
This indicates that Ti in the oxide matrix represented by this interface is in
a metallic state with valence intermediate between 3+ and 4+.
Thus Ti$^{4+}$ oxides with small band gaps
intermediate between the large gaps of materials
such as BaTiO$_3$ and the zero gap metal of the LaAlO$_3$ / SrTiO$_3$
interface could exist.
In general the charge balance, in other words the ionic states, of
atoms in solids are determined by the Ewald potential.
The key questions are whether it is possible to use this in realizable
layered structures to destabilize Ti$^{4+}$ sufficiently to lower
the band gap, without crossing over to Ti$^{3+}$ or a metallic state,
and if so, whether the resulting material can be made ferroelectric.

Bi$_4$Ti$_3$O$_{12}$
is an interesting starting material from this point
of view.
It has a polarization of $P \sim 50 \mu$C/cm$^2$ at room temperature,
\cite{cummins2,cummins,sawaguchi}
and it has a moderately low absorption edge of 3 eV.
\cite{oliveira,singh-bi4}
It has a layered structure based on a stacking
of perovskite BiTiO$_3$ and fluorite-like bismuth oxide  blocks (see below) and
can be grown in very high quality thin film form. \cite{lee}
The presence of separated oxide blocks in a material
amenable to thin film growth allows more possibilities for
chemical modification, while retaining the ferroelectric function.
\cite{armstrong,snedden,lee,park,chon}
According to recent first principles calculations,
the reduced band gap in this material arises 
from an imbalance between the Bi-O and perovskite
parts of the unit cell. \cite{singh-bi4}
It is known that alternate layerings based on Bi$_4$Ti$_3$O$_{12}$
can be grown in thin films. For example, Nakashima and co-workers recently
reported synthesis and properties of Bi$_5$Ti$_3$FeO$_{15}$.
Importantly, the polarization direction is almost perpendicular
to the layer stacking direction in Bi$_4$Ti$_3$O$_{12}$.
Here we exploit these facts to propose a new
ferroelectric material based on Bi$_4$Ti$_3$O$_{12}$ but with a lower
band gap of $\sim$1.8 eV, which is a much better value 
for exploiting the solar spectrum than typical oxide ferroelectric band gaps
of 3 eV or higher.
The proposed
material is based on an alternate layering with an extra perovskite
layer and an extra compensating Bi-O fluorite layer, to yield a 
formula Bi$_6$Ti$_4$O$_{17}$.
This amounts to adding an extra perovskite layer to the perovskite
block in Bi$_4$Ti$_3$O$_{12}$, which adds positive charge to this
block and compensating this by negative charge in an extra BiO$_2$
fluorite type layer.

Importantly, Ti$^{4+}$ contains no occupied $d$ orbitals, and
is non-magnetic. While this precludes any magnetic functionality,
it does offer some advantages: the lack of magnetic moments is favorable
for achieving reasonable mobility because there will not be strong
magnetic scattering, and the band gap will be of charge
transfer character and therefore associated
with strong optical absorption in contrast to gaps of $d$-$d$ character.

We performed full density functional
structural relaxations for Bi$_6$Ti$_4$O$_{17}$
and find a ferroelectric
structure. The calculated polarization is smaller than Bi$_4$Ti$_3$O$_{12}$,
but still sizable.
We then did calculations of the electronic structure using the
functional of Tran and Blaha, \cite{mbj} which is a modified Becke-Johnson
form that includes the kinetic energy density and yields greatly improved
band gaps for simple oxides and semiconductors.
\cite{mbj,singh-1,singh-2,kim}
Importantly, application
of this functional, which we denote TB-mBJ,
to Bi$_4$Ti$_3$O$_{12}$ yields very close
agreement with the measured experimental optical spectrum. \cite{singh-bi4}
The calculated band gap of the
new Bi$_6$Ti$_4$O$_{17}$ ferroelectric phase with the TB-mBJ functional
is 1.83 eV.

\begin{figure}
\includegraphics[width=\columnwidth]{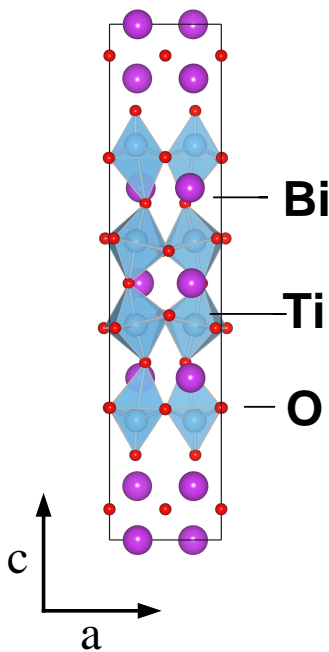}
\caption{(color online) Relaxed structure of Bi$_6$Ti$_4$O$_{17}$.}
\label{struct}
\end{figure}

As mentioned, we started with a full structural relaxation. This was
done using the generalized gradient approximation (GGA) of Perdew, Burke and
Ernzerhof (PBE). \cite{pbe}
We used the VASP package \cite{vasp1,vasp2}
with projector augmented wave (PAW) pseudopotentials, \cite{paw1,paw2}
for the initial structure relaxation.
This was done in a scalar relativistic approximation
with a plane-wave cutoff of 500 eV for the basis set and
with a Brillouin zone sampling based on a 4x4x1 {\bf k}-point mesh,
where $c$ is the long axis of the cell.
Energy and force convergence was tested.

\begin{table}
\caption{Internal coordinates of monoclinic $Pb$
(No. 7, unique axis a) Bi$_{6}$Ti$_4$O$_{17}$.
There are two formula units per cell, with all atoms occurring on general
sites. The lattice parameters are
$a$=5.479\AA, $b$=5.510\AA, $c$=25.278\AA, $\alpha$=90.965$^\circ$.}
\begin{tabular}{lccc}
\hline \hline
  & ~~~~~~~$x$~~~~~~~ & ~~~~~~~$y$~~~~~~~ & ~~~~~~~$z$~~~~~~~ \\
\hline
Bi   &    0.2481       &     0.0000     &       0.9997 \\
Bi   &    0.2520       &     0.5045     &       0.1043 \\
Bi   &    0.2254       &     0.5729     &       0.3143 \\
Bi   &    0.2349       &     0.5433     &       0.4981 \\
Bi   &    0.2199       &     0.4305     &       0.6830 \\
Bi   &    0.2518       &     0.4947     &       0.8949 \\
Ti   &    0.2640       &     0.9686     &       0.7655 \\
Ti   &    0.2607       &     0.0278     &       0.2315 \\
Ti   &    0.2637       &     0.0051     &       0.4169 \\
Ti   &    0.2616       &     0.0090     &       0.5796 \\
O    &    0.5005       &     0.2539     &       0.0596 \\
O    &    0.9996       &     0.2538     &       0.0596 \\
O    &    0.2678       &     0.0274     &       0.1648 \\
O    &    0.9989       &     0.2662     &       0.2535 \\
O    &    0.5058       &     0.2706     &       0.2563 \\
O    &    0.1801       &     0.9699     &       0.3439 \\
O    &    0.5499       &     0.1785     &       0.4108 \\
O    &    0.0356       &     0.2491     &       0.4356 \\
O    &    0.3276       &     0.9584     &       0.4976 \\
O    &    0.0302       &     0.2600     &       0.5606 \\
O    &    0.5383       &     0.1997     &       0.5847 \\
O    &    0.1779       &     0.0209     &       0.6529 \\
O    &    0.4939       &     0.2248     &       0.7409 \\
O    &    0.9985       &     0.2269     &       0.7424 \\
O    &    0.2673       &     0.9698     &       0.8320 \\
O    &    0.4996       &     0.2456     &       0.9392 \\
O    &    0.0002       &     0.2457     &       0.9392 \\
\hline
\end{tabular}
\label{tab-struct}
\end{table}

We started with an initial guess for the structure based on the layering of
Bi$_4$Ti$_3$O$_{12}$ with an additional perovskite BiTiO$_3$ block compensated
by an additional fluorite BiO$_2$ layer. We then fully relaxed this structure
for both the lattice parameters and internal coordinates
allowing monoclinic $Pb$ symmetry, which is the symmetry
of Bi$_4$Ti$_3$O$_{12}$. This was
continued until the stress tensor components were below 0.25 kBar and
the atomic forces were below 0.01 eV/\AA.
The final structure had lattice parameters
$a$=5.479 \AA, $b$=5.510 \AA, $c$=25.289 \AA, and monoclinic
angle $\alpha$=90.965$^\circ$. The resulting unit cell volume is
763.4 \AA$^3$.
We then performed a further relaxation of the internal coordinates
using the more precise all-electron general potential linearized
augmented planewave (LAPW) method, \cite{singh-book}
with no imposed symmetry.
These calculations were done with the WIEN2k code. \cite{wien}
We used a well converged LAPW basis with local orbitals to relax
linearization errors and include semicore states. \cite{singh-lo}
The resulting structure is depicted in Fig. \ref{struct} and the internal
coordinates are given in Table \ref{tab-struct}.
Importantly, the structure does not relax to a higher symmetry
non-polar group but stays essentially $Pb$.

The polarization was calculated using the Berry phase method \cite{berry3}
with the quantum espresso code.
\cite{qe}
The resulting polarization
is in the monoclinic plane close to the $b$-axis as defined
($c$=5.510 \AA), but rotated towards the long $c$-axis.
The calculated polarization is 19 $\mu$C/cm${^2}$ in the $b$-$c$
plane, 5.5$^\circ$ away from $b$.
It comes mainly from Ti and the Bi off-centering in the perovskite blocks.
Therefore the ferroelectricity can still exist in this charge imbalanced
structure.

\begin{figure}
\includegraphics[height=\columnwidth,angle=270]{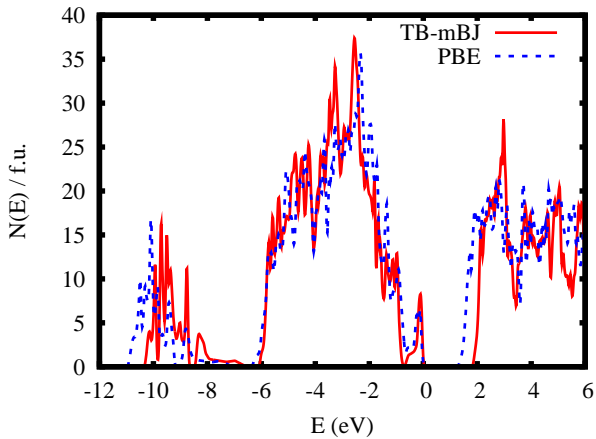}
\caption{(color online)
Electronic density of states of Bi$_6$Ti$_4$O$_{17}$ as obtained
with the PBE and TB-mBJ functionals based on the relaxed crystal
structure.
The states below $\sim$-7 eV are Bi $6s$ states, those from -7 eV to
the valence band edge at 0 eV are the O 2$p$ derived valence bands,
and those above 0 eV are the conduction bands.
}
\label{dos}
\end{figure}

We used the relaxed structure, as above, as input to electronic structure
calculations with the LAPW
method.
The calculations were performed relativistically, including spin-orbit,
with a 4x4x2 {\bf k}-point mesh for sampling the Brillouin zone.
Parallel calculations were performed with the TB-mBJ functional and the
standard PBE GGA.


\begin{figure}
\includegraphics[height=\columnwidth,angle=270]{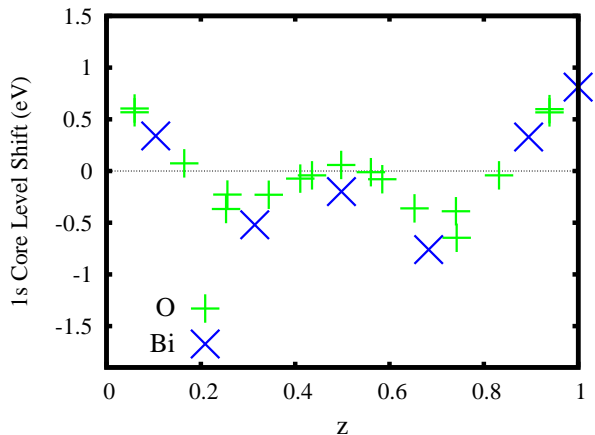}
\caption{(color online)
1$s$ core level positions of the O and Bi ions relative to the
average core level position
of the respective species as a function of fractional position $z$
along the $c$-axis as in Table \ref{tab-struct}.
}
\label{core}
\end{figure}

The resulting electronic densities of states are shown in Fig. \ref{dos}.
While the TB-mBJ band gap is larger than that obtained with the standard
PBE GGA, it is small compared to other oxide ferroelectric materials.
We obtain a value $E_g$=1.83 eV with this functional. For comparison,
the PBE value is 1.28 eV. Therefore this material is predicted to be
a low band gap oxide ferroelectric.
Turning to the nature of the gap, the valence bands are comprised mainly of
O $2p$ states, while the conduction bands are primarily from
Bi $6p$ and Ti $3d$ states, although as in other ferroelectrics based
on these elements, there is cross gap hybridization involving the
O $2p$ orbitals and the nominally unoccupied Bi $6p$
and Ti $3d$ states.
Thus the gap is of charge transfer character.
This is of importance because, unlike $d$-$d$
gaps, a charge transfer gap is generally associated with strong optical
absorption above the band edge.

The reduced gap and polarization relative to Bi$_4$Ti$_3$O$_{12}$ are
related.
As seen in Fig. \ref{struct} there are
large displacements relative to the O cages
of the ions away
from the center of the perovskite part of the unit cell especially
for the Ti and Bi ions away on the outer planes of the perovskite
block.
This is a consequence of the fact that the perovskite layers
have net positive charge, which is compensated by the Bi-O
layers. Thus cations are pushed away from the center of the perovskite
block towards the Bi-O block.

Actually, it might seem that the net positive charge of perovskite
blocks based on Bi$^{+3}$Ti$^{+4}$O$_3^{-2}$ (net +1 per unit)
means that one cannot add such a layer.
However, it should be emphasized that fully three dimensional perovskites
based on Ti and trivalent ions, i.e. LaTiO$_3$ and YTiO$_3$ for example,
do exist in nature. In these compounds Ti takes a trivalent +3 state.
What we are really doing is driving the Ti in the direction of a +3 state.
Importantly, stable Ti$^{4+}$ oxides typically
have band gaps of 3 eV or higher, while the present compound has a low band
gap and while in itself is an indication of chemical instability,
prior results for metallic conduction at Ti interfaces with charge
imbalance suggest that it may be possible to stabilize the present
stacking even though it has a small band gap.

The net positive charge in the perovskite layers is also the
origin of the reduced gap. The net positive charge pulls down
the energies of the cation orbitals in this region reducing the gap.
This effect can be clearly seen in the core level positions.
Both Bi and O ions occur in both the perovskite and Bi-O parts of the
unit cell.
Fig. \ref{core} shows the 1$s$ core level positions of these ions
as a function of the fractional coordinate along the $a$-axis relative
to the average core level position of the respective ion.
The perovskite part of the cell is approximately from
$x$=0.25 to $x$=0.75.
The deep 1$s$ levels do not participate in bonding, and therefore
their positions reflect the Coulomb potential.
As may be seen, the core level is pulled to higher binding energy in
the perovskite part of the cell, reflecting the positive charge in this
region (electrons are negatively charged, so electronic states
are pulled to higher binding energy by positive charge).
The variation in the core level positions is greater than
1 eV, which is enough
to explain the band gap reduction.
Also, if the compound is synthesized,
O $1s$ core level variations of this magnitude should be measurable.

Turning to the polarization, the local electric field induced cation
displacement reflects a rotation of the polarizations of the individual
perovskite layers to yield components along $a$ that mostly cancel between
the layers above and below the center.
In other words the cations in the outermost layers of the perovskite blocks
are pushed away from the center along the $a$-axis, reducing the mainly
$c$-axis net polarization.
Nonetheless, while
reduced relative to Bi$_4$Ti$_3$O$_{12}$, a sizable net
polarization remains.

To our knowledge the proposed compound Bi$_6$Ti$_4$O$_{17}$ has
not been reported. As such, a key question is how it could be made.
It does not appear in existing phase diagrams, and we note that there
are competing phases.
The basic idea underlying this work is to exploit constraints imposed
by layering, in this case charge imbalance. In a sense this is similar to
past work using epitaxial constraints to impose strain on various layers
in thin films to produce new ferroelectrics.
By analogy the most likely route would seem to be based
on thin film growth. In particular, as layer by layer growth of the
Bi-Ti-O system is perfected it may be become
possible to grow alternate stackings, including the one proposed in a
controlled way.

Even if it is not possible to grow the proposed material, it may be
possible to grow other stacking sequences using charge imbalance
to manipulate the band gap, starting
with ferroelectrics such as PbTiO$_3$ or (Ba,Sr)TiO$_3$.
For example, one could perhaps make superlattices consisting of
PbTiO$_3$ with alternating substitutions of TiO$_2$
layers by e.g. ScO$_2$ or AlO$_2$, which would have net negative
charge, and PbO/BaO by BiO or LaO, which would have net positive charge;
in this case the band gap would be controlled by the separation of the
positive and negative layers in the superlattice.
Furthermore, such a system may be ferroelectric if
one can keep the polarization perpendicular to the
layer stacking direction, e.g. by epitaxial strain as might be
obtained in this case
by growing on a rectangular (100) face of (001) poled tetragonal
PbTiO$_3$.

The key result of the present study is that we show computationally
by a specific example
that it is possible to have charge imbalanced titanates that combine
a low band gap with ferroelectricity.


Work at NUS was supported by the Singapore Agency for Science,
Technology and Research (A*STAR) through grant number 0721330044.
Work at ORNL was supported by the Department of Energy,
Office of Basic Energy Sciences, Materials Sciences and Engineering
Division (ferroelectricity) and
the Laboratory Directed Research and Development
Program of Oak Ridge National Laboratory, managed by UT-Battelle, LLC for
the U.S. Department of Energy (electronic structure).

\bibliography{bi}

\end{document}